\documentclass[fleqn,12pt,twoside]{article}
\usepackage{espcrc1}

\usepackage{graphicx}
\usepackage[figuresright]{rotating}

\newcommand{\DI}{\Delta I}
\newcommand{\Lam}{\Lambda}
\newcommand{\Sig}{\Sigma}
\newcommand{\Gn}{\Gamma_{nn}}
\newcommand{\Gp}{\Gamma_{pn}}

\newcommand{\AmS}{{\protect\the\textfont2
  A\kern-.1667em\lower.5ex\hbox{M}\kern-.125emS}}
  
\def\JLone<#1,#2>{#1}
\def\JLtwo<#1,#2,#3>{#2}
\def\JLyear<#1,#2,#3,#4>{#3}
\def\JLpage<#1,#2,#3,#4>{#4}
\newcommand\JL[1]{\JLone<#1>\ {\bfseries \JLtwo<#1>} (\JLyear<#1>) \JLpage<#1>}
\def\Jpage<#1,#2,#3>{#3}

\newcommand\andvol[1]{{\bfseries \JLone<#1>} (\JLtwo<#1>) \Jpage<#1>}

\newcommand\PTP[1]{Prog.\ Theor.\ Phys.\ \andvol{#1}}

\newcommand\PR[1]{Phys.\ Rev.\ \andvol{#1}}

\newcommand\PRC[1]{Phys.\ Rev.\ C\ \andvol{#1}}
\newcommand\PRD[1]{Phys.\ Rev.\ D\ \andvol{#1}}

\newcommand\PRL[1]{Phys.\ Rev.\ Lett.\ \andvol{#1}}
\newcommand\PL[1]{Phys.\ Lett.\ \andvol{#1}}

\newcommand\PLB[1]{Phys.\ Lett.\ B\ \andvol{#1}}
\newcommand\NP[1]{Nucl.\ Phys.\ \andvol{#1}}

\newcommand\IJMP[1]{Int.\ J.~Mod.\ Phys.\ \andvol{#1}}

\title{Models of the Nonmesonic Weak Decay}

\author{Makoto Oka\address[Titech]{Department of Physics, Tokyo Institute of Technology,
        Meguro, Tokyo 152-8551, Japan}%
        \thanks{email: oka@th.phys.titech.ac.jp}
		}
       
\begin{document}

\maketitle

\begin{abstract}
I review the current status of understanding the mechanism of the nonmesonic weak decays (NMWD)
of hypernuclei. Long standing problem on the $\Gn/\Gp$ ratio has been solved by considering
short-range weak interactions properly. This leaves a few remaining problems, such as
the asymmetry of the emitted proton with respect to the $\Lambda$ 
polarization and validity of the $\DI=1/2$ in the nonmesonic weak decay. 
\end{abstract}

\section{INTRODUCTION}

The year 2003 marks the 50th anniversary of the discovery of hypernucleus.
In these 50 years, the physics of hypernucleus and strange hadrons has played a key role in
understanding both the strong and weak interactions of quarks (and gluons).
Starting from the eight-fold way, or the flavor SU(3) symmetry, various symmetries of 
hadron spectroscopy and interactions, such as current algebra, were born from the strangeness physics. 
In 1990's, we have seen a marvelous development both in experimental and theoretical hypernuclear physics.
It has accomplished fairly good description of the strong, electromagnetic and weak interactions of hyperons 
in nuclear medium. Especially a series of new 
experimental data on the weak decays have lead us to a nearly full understanding of the mechanisms
of the weak decays of the hyperon in nuclei.  This is the subject of this report.

In section 2, I summarize the current status of various theories and models of the nonmesonic weak decay.
In section 3, the model calculations are compared with recent experimental data.
In section 4, I concentrate on the $\DI=1/2$ rule and argue that the rule may not be valid in the NMWD.
In section 5, conclusions are given.

\section{MODELS}

The weak decay of $\Lam$ in hypernuclei is dominated by the nonmesonic decay modes,
$\Lam N \to NN$, because the mesonic decay, $\Lam\to N\pi$ is suppressed.  This is 
easily understood by realizing that the mesonic decay emits a nucleon of momentum,
$\sim 100$ MeV/c, and thus is suppressed by the Pauli exclusion in the final states. 
In contrast, the nonmesonic weak decay (NMWD) may emit nucleons with momentum $\sim 400$ MeV/c.
This momentum corresponds to the region of the $NN$ interaction where the short-range
repulsion is important.  It indicates that the NMWD of $\Lam$ hypernuclei 
probes the short distance part of the baryonic weak interaction.

The nonmesonic decay of $\Lam$ is most simply realized by two kinds of the two-body decay modes,
$\Lam p \to pn$ (proton induced) and $\Lam n \to nn$ (neutron induced), while more complicated
processes, like $\Lam N N \to NNN$, may contribute.  
We start with the mechanism of the two-body decay modes.

\subsection{One pion exchange (OPE)}
Study of the hypernuclear weak decays showed steady but significant
development in 1990's.  The progress was mostly lead by the new generation of experiments, carried out
with high intensity secondary beams of mesons combined with high resolution spectrometers.
New developments include measurement of the decay neutron spectrum and also back-to-back
coincidence measurement.

Before such progress, our knowledge on NMWD was very limited. 
The simplest view was proposed just after the discovery of the hypernucleus by
Ruderman and Karplus\cite{RK56}, where the authors considered one-pion exchange
process with a $\Lam N\pi$ weak vertex for the NMWD and 
calculated the ratio of the mesonic to nonmesonic decay rates in nuclear matter.
They concluded that the then-available emulsion data are consistent only if the spin of $\Lam$ is
${1\over 2}$ or ${3\over 2}$.
Later, Block and Dalitz\cite{BD63} compared the one-pion exchange (OPE) process
with experiment carefully.  
The weak coupling constants at the $\Lam N\pi$ vertex are fixed by the decay 
rate and the asymmetry of the $\Lam\to N\pi$ decay.
It was realized that although the total decay rate is roughly reproduced the ratio of
the neutron induced decay rate, $\Gn$, and the proton induced decay rate, $\Gp$,
is in trouble.  In fact, the large momentum transfer enhances
the contribution of the tensor part of OPE, which induces a strong 
$\Lam p (^3{\rm S}_1) \to np (^3{\rm D}_1)$ transitions. 
Because of the isospin symmetry in the final state, no tensor transition is allowed for the 
$\Lam n \to nn$, which leads to the relation $\Gp \gg \Gn$.
Unfortunately, the experimental data do not show this disparity between the two modes.

\begin{figure}[htbp]
\begin{center}
\includegraphics[height=6cm]{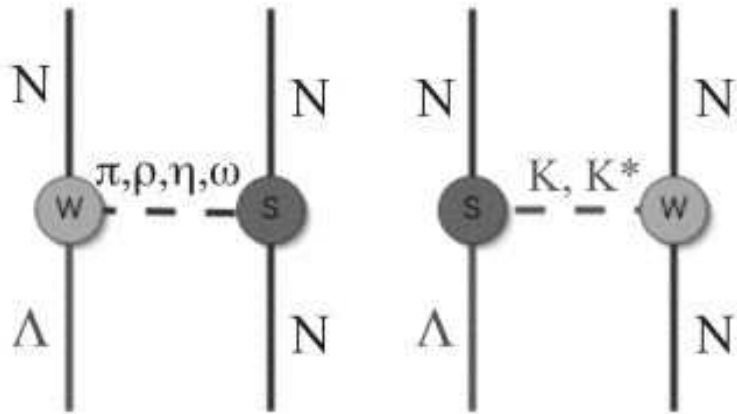}
\caption{{One meson exchange weak process.}}
\label{Fig1}
\end{center}
\end{figure}

\begin{figure}[htbp]
\begin{center}
\includegraphics[height=6cm]{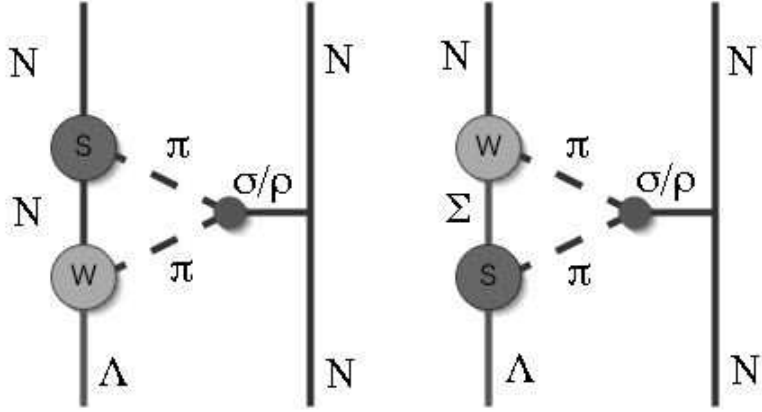}
\caption{Correlated two pion exchange in NMWD.}
\label{Fig2}
\end{center}
\end{figure}
\subsection{Heavy meson exchanges}
As the transferred momentum is high between two baryons, we expect to have shorter range
interactions contribute to the NMWD.  
It is natural to follow the description of the strong interaction of the baryons, in which the
various heavy meson exchanges are coherently added up. 
The $\rho$ exchange for the weak decay was first introduced by McKellar and Gibson\cite{BM84}. 
Takeuchi et al.\cite{TTB85} applied their model 
to $A=4$ and 5 hypernuclei.  
These two pioneering works considered only the tensor part of the $\rho$ exchange, where 
the ``factorization" ansatz was used to determine the weak $\rho\Lam N$
coupling constants.

Later, Dubach et al.\cite{DFHD96} proposed to employ the SU(3)/SU(6) symmetry to
evaluate the weak coupling constants 
of the pseudoscalar and vector mesons to the octet baryons. The parity violating (PV) couplings of the
pseudoscalar mesons are determined by the SU(3) symmetry from the $\Lam (\Sig) \to N\pi$ couplings, 
while the SU(6) symmetry
is required for the vector mesons and furthermore an extra parameter is introduced.
The parity conserving (PC) part of the weak coupling is assumed to come from the baryon pole
diagrams.  Similar assumptions have been employed in the calculation of nonleptonic decay
of the hyperons.  It is known, however, that they do not reproduce both the PC and PV decay
rates satisfactorily.  Thus it is important to evaluate the weak coupling constants directly
from QCD.

Dubach et al.\cite{DFHD96} analyzed the NMWD of $\Lam$ in nuclear matter by employing the above
prescription for the weak coupling constants.
All the members of the pseudoscalar and vector nonet mesons are taken into account.
The full vector meson exchanges including the central part of the interaction 
were considered also by Parreno et al.\cite{PRB97} for
the calculation of the decay rates of $^{12}_{\Lam}{\rm C}$.

It is well known that exchanges of two pions correlated into the scalar-isoscalar meson, ``$\sigma$'',
is crucial in nuclear force to make nuclear bound states.
In the weak decay, Shmatikov\cite{Shmatikov94} and Itonaga et al.\cite{Itonaga94} 
considered $2\pi$ in the scalar $\sigma$ channel and $\rho$ channel.
The advantage of this approach is that the weak coupling constant can be estimated  from
the weak $\Lam\to N\pi$ and $\Sig\to N\pi$ vertices.
The contributions are found to be sizable in the $J=0$ transitions and
also in the tensor transition, although the results from the two groups 
do not necessarily agree with each other.

\subsection{Direct quark}
As the short range baryonic interaction plays important roles, the nucleon
substructure must be taken into account for the complete understanding of NMWD.
The $A$ dependence of the hypernuclear life-times also indicates that 
short-range interaction plays an important role in NMWD.\cite{KEKE307}
It is known that the short range $NN$ repulsion comes from the quark exchange
interaction.\cite{OY80}
Similarly, the direct quark (DQ) weak process was introduced in order to explain 
the short-range part of the NMWD of hypernuclei.
 
\begin{figure}[htb]
\begin{center}
\includegraphics[height=6cm]{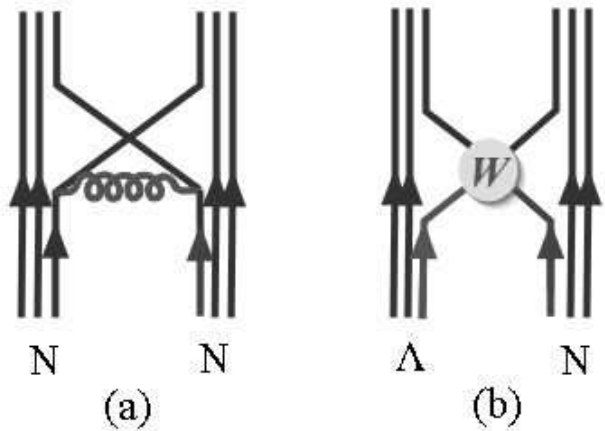}
\caption{(a) Quark exchange diagram for nuclear force and (b) direct quark (DQ) diagram for NMWD.}
\label{Fig3}
\end{center}
\end{figure}
In the DQ calculation, an effective hamiltonian is introduced for the weak interactions
of quarks, which consists of various four-quark vertices, $s u \to d u$ and $s d \to d d$.
In the pioneering work done by Cheung et al.\cite{CHK83}, these operators were not correctly
taken into account. Maltman and Shumatikov\cite{MS94} considered the $\pi$ and $K$ exchange with
DQ, and pointed out that the $\DI=1/2$ may be violated by the DQ.
Inoue et al.\cite{ITO94} showed that the DQ $+\pi$ exchange gives a large $\Gn/\Gp$ 
ratio, which is one of the difficulties of OPE. They also pointed out that 
a strong $\DI=3/2$ transition is realized in the $J=0$ amplitudes.
Later, Sasaki et al.\cite{SIO00} showed that the $\Gn/\Gp$ ratio becomes as large as
$0.5 \sim 1$ when we take account of $\pi$, $K$ and DQ exchanges.

\subsection{Other theoretical studies}
Another notable approach to the baryonic weak interaction is 
the effective interaction approach, which was first considered by Block and Dalitz\cite{BD63}, 
and later elaborated by Jun et al.\cite{Jun98} and Parreno et al.\cite{PBH03}
They are mostly phenomenological but the last one is based on the chiral effective theory.
The latter can reproduce the basic features of the experimental data employing 
sufficient number of parameters.

Nuclear effects on the NMWD are considered in various aspects.
The medium effects on OPE and one-boson exchange (OBE) have been studied by many groups, 
where the short-range correlation and final state interaction are two important effects.\cite{OS85}
Contribution of two-nucleon induced decay, $\Lam NN\to NNN$, especially in extracting the
$\Gn/\Gp$ ratio was studied by Spanish groups.\cite{ROS94}  Their results seem to agree 
rather well with recent experimental spectrum.

\begin{figure}[htbp]
\begin{center}
\includegraphics[height=6cm]{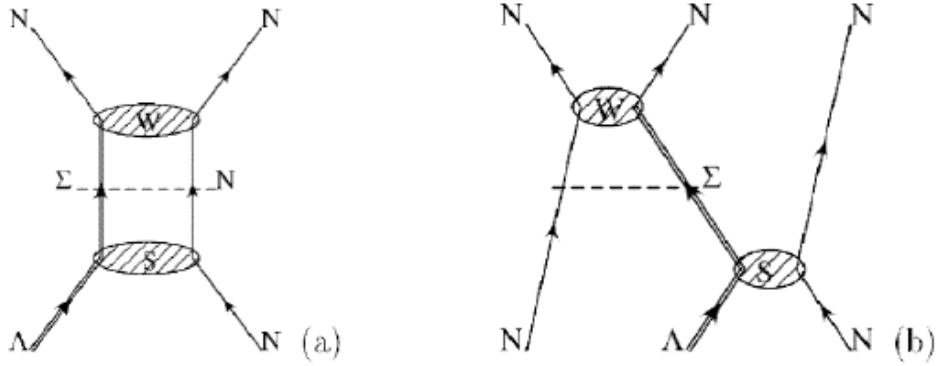}
\caption{Virtual $\Sigma$ contributions. (b) shows a coherent process.}
\label{Fig4}
\end{center}
\end{figure}

The roles of virtual $\Sig$ excitation is another interesting contribution to the NMWD.
It was first considered by Bando et al.\cite{BST88}, who introduced the $\Lam N \to \Sig N \to NN$
decay process.
Recently, Sasaki et al.\cite{SIO02} pointed out the importance of 
coherent $\Sig$ mixing in the $A=4$ ($I\ne 0$) hypernuclei,
and found that it gives 10\% effects.  It may be closely related to the ``overbinding'' problem of the S shell
hypernuclei\cite{NA} and is an interesting subject to pursue further.

\section{EXP v.s. THEORY}

Main observables in NMWD are the nonmesonic decay rates, $\Gamma_{\rm NM}$, 
the partial rates $\Gp$, $\Gn$ and 
the proton asymmetry parameter $\alpha_p({\rm NM})$. 

\begin{table}[htdp]
\caption{Current status of the comparison of data and theory.}
\begin{center}
\begin{tabular}{|c|c|c|c|c|}
\hline
$^5_{\Lam}{\rm He}$&$\Gamma_{\rm NM}$&$\Gn /\Gp$&$\alpha_p({\rm NM})$&\cr
\hline
$\pi+K+{\rm DQ}$&0.52&0.70&$- 0.68$&Sasaki et al.\cite{SIO00}\cr
OBE (all)&0.32&0.46&$- 0.68$& Parreno et al.\cite{PRB97}\cr	
$\pi+K+\omega+2\pi/\rho, \sigma$&0.42&0.39&$-0.33$ or 0.12&Itonaga et al.\cite{Itonaga02}\cr
\hline
Exp&$0.41\pm0.14$&$0.44\pm0.11$&$0.09\pm 0.08$&KEK E462/E508\cite{KEKE462}\cr
\hline
\end{tabular}
\end{center}
\label{Table1}
\end{table}

The theoretical prediction of the $n/p$ ratio, $\Gn /\Gp$, has a long history.
In 1963, Block and Dalitz showed that the OPE underestimates the ratio by a factor 10.
Until 1997, the OBE with all mesons showed no improvement giving about $\Gn /\Gp \sim 0.1$.
The calculation in the $\pi+{\rm DQ}$ model performed in 1998 gave a larger value, 0.49, but
adding the $K$ exchange at that time reduced the value to 0.20 for $^5_{\Lam}{\rm He}$.
In 2000, we found an error in the sign in the $K$ exchange amplitudes commonly used.
The revised values of $f_p(K)$ and $f_n(K)$, which denote the $^3S_1\to ^3P_1$ ($I =0$) 
transition, were shown to give a larger value of the $n/p$ ratio,  
0.45 for $\pi+K$ and 0.70 for $\pi+K+{\rm DQ}$ 	for $^5_{\Lam}{\rm He}$.\cite{SIO00}
The meson exchange calculations have also agreed with this conclusion. 

Now the KEK E462 and E508 experiments\cite{KEKE462} 
have successfully singled out the back-to-back
decay events, which are not contaminated by the final state rescattering, and have
obtained the relatively lower value of the $n/p$ ratio.  
Finally, the theories and experiments agree with each other and the $n/p$ puzzle has been
solved.

The nucleon spectrum emitted from the NMWD has also been measured and compared with
the theoretical one.  Agreement is not complete yet and it seems that we still need more
realistic final state description.

The remaining problem is in the proton asymmetry parameter, $\alpha_p$, which 
describes the decay proton asymmetry with respect to the polarization of $\Lam$.\cite{NOSK99}
As is shown in Table 1, theoretical calculations mostly give a large negative value, while recent
experiments indicate small or positive asymmetry.

Future studies of NMWD may concentrate on the $J =0$ decay amplitudes,
which have not been determined well experimentally nor theoretically.
They are the key to understand the proton asymmetry	$\alpha_p$ and
$\DI =3/2$ in NMWD.

More observables are being measured in the future.  One of the interesting experiments 
is the reversed process, i.e., the weak production of $\Lam$ in $np \to \Lam p$ scattering.
It was shown that the $J=1$ amplitudes may dominate the process.\cite{Lprod}
Although the expected cross section is tiny, 
of the order $\sigma\sim 10^{-39}\sim 10^{-40} {\rm cm}^2$,
an experiment is planned at RCNP\cite{Minami}.

The decay of  double hypernuclei is a new and interesting subject, as this decay has a
new decay modes, 
$\Lam\Lam \to \Lam N$ or $\to\Sig N$, which are supposed to be dominated 
by the $J =0$ amplitudes.\cite{LLdecay}

\section{$\DI=1/2$ dominance in the nonleptonic strangeness decay}

It is well known that the strangeness changing weak decay strongly favors
the $\DI=1/2$ transition.  The experimental ratios of the $\DI=1/2$ and
$\DI=3/2$ amount to 20 in the decay amplitudes of the kaon and the hyperon. 
Yet, the mechanism for the
$\DI=1/2$ dominance is not completely understood.  Indeed, the standard
theory of the electroweak interaction predicts the amplitudes of the same
order, which can be seen from the fundamental vertex,
$s \to u+W^- $, $W^-\to d+ \bar u$ ($I=1$).
The final $ud\bar u$ system has either isospin 1/2 or 3/2 with the ratio 2 to 1,
which comes from the Clebsch-Gordan coefficients.
Thus naive expectation contradicts with data.
\begin{figure}[htbp]
\begin{center}
\includegraphics[width=15cm]{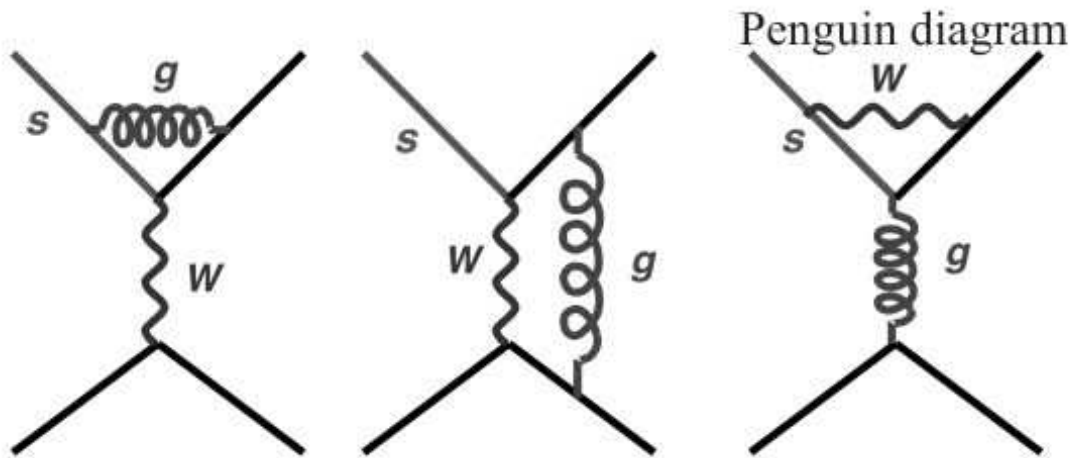}
\caption{QCD corrections for the $\Delta S=1$ nonleptonic weak interaction.}
\label{Fig5}
\end{center}
\end{figure}

The difficulty is partially solved by considering the perturbative QCD corrections
on the weak transition process.  The one-loop QCD corrections associated with
the renormalization group (RG) improvement are shown to enhance (suppress) 
the $\DI=1/2$ ($\DI=3/2$) amplitudes.\cite{GL74,AM74,VZS77,GW79}
The mechanism can be understood by considering the color-flavor structure of
the gluon exchange interactions between quarks.  The color-magnetic attraction in 
the scalar diquark, $I =0$, $S =0$, $C =\bar 3$ $ud$,  in the final state enhances 
the $\DI=1/2$.  Also the so-called Penguin diagram, which is purely $\DI=1/2$ and
mixes the right-handed flavor-singlet current, was shown to contribute significantly.\cite{VZS77}

The low energy effective interaction for the nonleptonic $s$ decay is given in terms of
various four quark operators, whose coefficients are given as the solution of the 
RG equation at a renormalization scale of order 1 GeV.
The RG equation takes into account the leading log summation and also the operator
mixings. The coefficients show clearly the enhancement of the $\DI=1/2$ and
the suppression of the $\DI=3/2$ vertices.

Unfortunately, the perturbative correction seems not sufficient in explaining the
observed $\DI=1/2$ to $\DI=3/2$ ratio. The consensus is that extra 3-5 times
enhancement is acquired from ``nonperturbative'' QCD corrections.
This is formally given by the corrections to the matrix elements of the operators, which are
renormalized at a scale $\mu\sim 1$ GeV, as all the physics above $\mu$ should be 
included in the coefficients.  The matrix elements are to be evaluated by, for instance,
the lattice QCD, effective theories, and models.  

As for the $\DI=1/2$ enhancement in the $K$ decay, Morozumi et al.\cite{MLS90} proposed 
$K\to\sigma\to 2\pi$ mechanism, in which $\sigma$ is
a scalar $I=0$ meson of mass $\sim 600$ MeV.
This process enhances the $\DI=1/2$ amplitude as the $\sigma$ mass is close to the kaon
mass.
A model calculation based on the NJL model was performed and showed that the
experimental ratio can be reproduced fairly well in this mechanism.\cite{ITO95}

When the hyperon decays are concerned, the $\DI=1/2$ enhancement may be
understood by combination of the soft-pion relation and the color symmetry 
argument.  Miura-Minamikawa\cite{MM} and Pati-Woo\cite{PW} (MMPW) showed that no $\DI=3/2$ amplitude
comes from the weak vertex diagram in which either the initial two quarks or the final two quarks 
belong to the same baryon. This theorem is a consequence of the color symmetry 
of the constituent quarks of the baryon.  For the pionic decays of hyperons, the PV
transition matrix elements can be reduced into the baryon-baryon matrix elements
by the soft-pion technique, for example,
\begin{equation}
 \langle n\pi^0(q) | H^{\rm PV} | \Lambda\rangle \to -{i\over f_{\pi}}
\langle n| [Q_5^0, H^{\rm PV}]| \Lambda\rangle 
= -{i\over 2f_{\pi} } \langle n| H^{\rm PC} | \Lambda\rangle
\end{equation}
where we use the following relations satisfied by the weak effective Lagrangian,
\begin{equation}
\, [Q_R^a, H_W] = 0 \qquad \, [Q_5 ^a, H^{\rm PV}] = - [I^a, H^{\rm PC}] .
\end{equation}
Then the baryon-baryon matrix elements, ex. 
$\langle n| H^{\rm PC} | \Lambda\rangle$, evaluated in the constituent quark model
contains only $\DI=1/2$ transitions due to the MMPW theorem.
As for the PC amplitudes, contributions of the pole diagrams again guarantee
the $\DI=1/2$ dominance.

Are the the baryon-baryon weak matrix elements large enough to explain baryon decays?
The matrix elements are sensitive to the baryon size.
In the DQ calculation, the nonrelativistic quark model wave function with $b = 0.5$ fm is used.  
This gives a large enough matrix element for the $\Lam\to N\pi$ decay.
The MIT bag model tends to underestimate the matrix elements.

The next interesting question then is whether the $\DI=1/2$ is dominant in NMWD
of the hyperons.
In the two-baryon weak interactions, we do not see any enhancement mechanism 
for $\DI =1/2$, nor suppression of $\DI =3/2$.  Indeed, it has been pointed out that
the DQ mechanism gives $\DI=3/2$ amplitudes comparable to $\DI=1/2$.\cite{MS94,ITO94}
Maltman and Shmatikov also suggested that the $\Lam, \Sig \to N \rho$ couplings 
may contain significant $\DI =3/2$.\cite{MS95}
It is therefore important to check whether the $\DI=3/2$ transition is significant in
the hypernuclear decays, for which a few proposal were made for the decays of the
$S$ -shell hypernuclei $A=4$ and 5.\cite{D87}

\def\half{1/2}
The $\DI$ of the weak transition will be clearly seen in the $J=0$ decay amplitudes.
We define the ratio,
$x= {\Gn^0/\Gp^0}$,
where $\Gp^J$ ($\Gn^J$) denotes the proton induced decay rate for the total
angular momentum $J$ (= 0 or 1). 
Then $x$ is a clear indicator of $\DI$:
\begin{equation}
 x=\cases{2& for $\Delta I= \half$\cr
\half &for $\Delta I= 3/2$\cr}
\end{equation}

The partial decay rates of $^4_{\Lam}{\rm He}$, $^4_{\Lam}{\rm H}$
and $^5_{\Lam}{\rm He}$, are the key quantities.
We define two ratios of the observables:
\begin{eqnarray}
\alpha &\equiv& {\Gamma_{NM} (^4_{\Lambda}H) \over \Gamma_{NM} (^4_{\Lambda}He) } 
\qquad
\beta \equiv{\Gamma_{nn} (^5_{\Lambda}He) \over \Gamma_{pn} (^5_{\Lambda}He) }
\end{eqnarray}
Then the following theorem can be proved if one assumes that the nonmesonic decay is
dominated by the two-body processes:\\

\smallskip
{\bf Theorem:  If $\alpha > \beta$  then  $x  < 1/\alpha$.} 
\medskip

\noindent The current experimental data are not conclusive:
$$ \alpha ={0.17\pm 0.11\over 0.17 \pm 0.05} \qquad \beta = 0.48\pm 0.10$$
but it indicates $x<1$ and therefore possibility of violation of the $\DI=1/2$ rule.

A related topic is the $\pi^+$ decay of hypernuclei, which is not induced by the 
$\Lam\to N\pi$ decay, but needs to involve another proton.  I applied the
soft pion technique again to the $\pi^+$ emission and showed that the decay through
$p\Lam \to n\Sig^+ \to nn\pi^+$ is strongly hindered in the soft pion ($S$ wave) limit.\cite{Oka99}
Instead, the $\DI =3/2$ transition gives nonvanishing matrix element in the soft pion limit,
and therefore the $\pi^+$ emission rate is directly connected to the $\DI=3/2$ amplitudes of NMWD.
Thus a large $\pi^+$ emission rate may indicate the violation of the $\DI=1/2$ rule in NMWD.

\section{Conclusion}

After 50 years of study, the NMWD of hypernuclei is fairy well understood 
in terms of the models of meson exchanges and also by taking account of the quark substructure of the baryon.
The long standing problem of the $\Gn/\Gp$ discrepancy has been solved in cooperation of
the theoretical and experimental efforts.
The decay nucleon spectrum indicates strong final state interactions.
Yet, we still need some refinement to reproduce the proton asymmetry parameter.
The predicted $\DI =1/2$ violation should be tested by experiment.
It is crucial to measure $\Gamma_{\rm NM} (^4_{\Lam}{\rm H})$ precisely.
The $\pi^+$ decay rate may also be useful in determining the role of $\DI=3/2$ component
in NMWD.

The roles of virtual $\Sig$ (and also $\Xi$ in the double hypernuclei) are left for the future  study.
After all, our goal is to understand the structure and dynamics of hypernuclei as well as to 
study the QCD corrections on the nonleptonic weak interaction in nuclei.
In 2007, the J-PARC is expected to start new series of hypernuclear experiments.
Combined studies of spectroscopy and weak decay may provide high quality data so that
high precision analyses become possible.

\end{document}